\documentclass[aps,prd,twocolumn,nofootinbib,showpacs,preprintnumbers,floatfix,longbibliography]{revtex4-2}
\usepackage{natbib}
\usepackage{longtable}
\usepackage[pdftex]{graphicx}
\usepackage[usenames]{color}
\usepackage[dvips]{epsfig}

\usepackage{appendix}
\usepackage{hyperref}
\usepackage{bbm}
\usepackage{marvosym}
\usepackage{verbatim}
\usepackage{psfrag}
\usepackage{slashed}
\usepackage{mathtools}
\usepackage{amsmath}
\usepackage{amssymb}
\usepackage{wasysym}
\usepackage{multirow}
\usepackage{subfigure}
\usepackage{soul}
\usepackage{diagbox}
\usepackage{siunitx} 
\usepackage{array} 
\usepackage{amssymb}  
\usepackage{booktabs,amsmath}
\usepackage{bbold}
\usepackage[table,x11names]{xcolor}
\definecolor{maroon}{cmyk}{0,0.87,0.68,0.32}
\usepackage{amsfonts}

\definecolor{boxcolor}{HTML}{ffe6a1}

\def \Nc {{N_{c}}}

\newcommand{\beqn} {\begin{equation}}
\newcommand{\eqn} {\end{equation}}

\def \beq{\begin{equation}}
\def \eeq{\end{equation}}
\def \bea{\begin{eqnarray}}
\def \eea{\end{eqnarray}}

\def \bet0{\beta_0}
\def \bet1{\beta_1}
\def \simgt{\,\rlap{\lower 7.5 pt\hbox{$\mathchar \sim$}}\raise 3 pt \hbox{$>$}\,}
\def \simlt{\,\rlap{\lower 7.5 pt\hbox{$\mathchar \sim$}}\raise 3 pt \hbox{$<$}\,}

\def\lsim{\raise0.3ex\hbox{$<$\kern-0.75em\raise-1.1ex\hbox{$\sim$}}}
\def\gsim{\raise0.3ex\hbox{$>$\kern-0.75em\raise-1.1ex\hbox{$\sim$}}}

\newcommand{\SU}{{\rm SU}}
\newcommand{\U}{{\rm U}}

\DeclareMathAlphabet{\mathpzc}{OT1}{pzc}{m} {it}
\begin{document}
\title{Quantum sampling on a quantum annealer for large volumes in the strong coupling limit for gauge group \U(3)}

\author{Jangho Kim$^{\rm a,b,c}$}
\email{jangho@physik.uni-bielefeld.de}
\author{Thomas Luu$^{\rm a}$}
\email{t.luu@fz-fuelich.de}
\author{Wolfgang Unger$^{\rm b}$}
\email{wunger@physik.uni-bielefeld.de}
\affiliation{$^{\rm a}$ Institute for Advanced Simulation (IAS-4) \& JARA-HPC, Forschungszentrum J\"ulich}
\affiliation{$^{\rm b}$ Fakult\"at f\"ur Physik, Bielefeld University, D-33615 Bielefeld, Germany}
\affiliation{$^{\rm c}$ Lattice Gauge Theory Research Center, Department of Physics and Astronomy, Seoul National University, Seoul 08826, Republic of Korea}

\begin{abstract}
In our previous studies~\cite{Kim:2023sie,Kim:2023pjk}, we confirmed that a quantum annealer can be used for importance sampling of gauge theories. In this paper, we extend the previous results to larger 2-dimensional and 4-dimensional lattices to generate ensembles for $\U(3)$ gauge theory in the strong coupling limit. We make use of the D-Wave quantum annealer to generate histograms for sub-lattices, and use the Metropolis-Hastings algorithm to determine thermodynamic observables and their dependence on the physical parameters on large volumes. We benchmark our results to those obtained from classical Monte Carlo simulations. 
\end{abstract}

\maketitle
\tableofcontents

\section{Introduction}
Quantum annealers are promising devices for ensemble generation in statistical physics and field theory. 
The examples of quantum sampling range from boson sampling~\cite{Lund:2017cfv} to sampling proteins~\cite{Ghamari:2024iat}. 
Also, many different methods are on the market for quantum annealers: the noisy Gibbs sampler~\cite{Vuffray:2020hjh}, methods to investigate phase transition~\cite{Wild:2021ene}, hybrid approach~\cite{Jattana:2024otq}, 
and the annealer as a quantum thermal sampler~\cite{Izquierdo:2020acy}.
Quantum sampling of statistical ensembles requires however an additional step to obtain ensembles according to the target equilibrium distribution that depends on the physical parameters of the investigated theory~\cite{Sandt:2023ewm,Ghamari:2022jyc,Shibukawa:2023vke}. 
In contrast to gate-based quantum computing, quantum annealers are restricted to optimization problems, and a crucial step is to map the physical system on the qubits of the annealer. This embedding poses a scalability challenge \cite{Quinton:2024rkl,Weinberg:2020mba}, which limits the usefulness for realistic applications further. 
Despite these challenges, some progress in high energy physics such as scalar field theory~\cite{Klco:2018zqz}, lattice gauge theory with non-Abelian gauge groups $\SU(2)$~\cite{ARahman:2021ktn}, dihedral group $D_3$~\cite{Fromm:2022vaj}, and for building blocks of standard model physics ~\cite{Illa:2022jqb} has been made. We address in this paper also a non-Abelian gauge group $\U(3)$, which at strong coupling has similar properties as $\SU(3)$ at vanishing quark density, as explained in Sec.~\ref{sec:model}.

On a classical computer, via a Monte Carlo, importance sampling is realized e.g.~by the Metropolis algorithm. The effective theory which we study is based on the dual formulation of lattice QCD at strong coupling, and is governed by a constraint on the dual variables, which restricts admissible configurations. Depending on the physical parameters, ensemble generation is a difficult task, specifically at low temperature where the entropy of the system grows rapidly. In contrast, on a quantum annealer, the admissible configurations can be generated easily with the so-called QUBO formalism which scales independently of the physical parameters such as the temperature, as explained in Sec. \ref{sec:model}.

The scope of this paper is to address larger volumes compared to our previous studies~\cite{Kim:2023sie,Kim:2023pjk}, which requires new strategies to take advantage of the quantum annealer.  We use the D-Wave advantage system with the Pegasus topology~\cite{dwave_pegasus_topology} that has 5760 physical qubits and is the most advanced annealer up to date. In Fig.~\ref{fig:topology} (left) we show its connectivity. This system can however only accommodate small lattice volumes.  Large volumes, on the other hand, require an alternative scheme that  samples smaller sub-lattices. The central idea of our work is to make use of an iterative scheme that embeds sub-lattices in parallel with specific boundaries that are fixed by the previous quantum sampling step.  This scheme ultimately allows us to generate larger volume configurations.

After short introduction about the effective theory in Sec.~\ref{sec:model}, we explain the technical aspects to address large volumes in Sec.~\ref{sec:tech}. We present our results on the volume and parameter dependence of observables in Sec.~\ref{sec:results}.

\section{Effective theory of Lattice QCD \label{sec:model}}
Our effective theory is derived from staggered fermions in the strong coupling limit, which is well known to have a dual representation in terms of integer-valued dual variables:
\begin{widetext}
\begin{align}
\label{eq:L}
S_F &=
\sum_{x,\nu} \frac{\gamma^{\delta_{\nu 0}}}{2}\eta_{\nu}(x)
\big(
    e^{\mu\delta_{\nu 0}} \overline{\chi}(x) U_{\nu}(x) \chi(x+\hat{\nu})
  - e^{-\mu \delta_{\nu 0}} \overline{\chi}(x+\hat{\nu})U^{\dagger}_{\nu}(x) \chi(x)
\big)+am_q\sum_x{\bar{\chi}_x\chi_x}
\end{align}
\begin{align}
\label{eq:dual}
Z&=\sum_{\{d,m\}}
\prod_{b=(n,\hat{\nu})}\frac{(N_c-d_\nu(n))!}{N_c!d_\nu(n)!}\gamma^{2d_\nu(n)\delta_{\hat{0},\hat{\nu}}}
\prod_{n}\frac{N_c!}{m(n)!}(2am_q)^{m(n)}\,,
\end{align}
\end{widetext}
where Eq.~\eqref{eq:dual} is obtained from Eq.~\eqref{eq:L} by integrating out the gauge variables first, and afterwards the Grassmann variables~\cite{Rossi:1984cv}. 
As we restrict to $\U(3)$ gauge group here, baryons are absent in Eq.~\eqref{eq:dual}, they would form self-avoiding world-lines for $\SU(3)$.
We have studied $\SU(3)$ in~\cite{Kim:2023pjk} but in this work we restrict ourselves to $\U(3)$ for simplicity. 
 The dual variables in Eq.~\eqref{eq:dual} consist of so-called monomers $m(n)$ and dimers $d_\nu(n)$ (with $n$ a lattice coordinate) 
Dimers correspond to pion hops, for $\Nc=3$  there can be up to 3 pions on the same link (due to the quark sub-structure), the monomers contribute to the chiral condensate. On each site, the number of monomers and dimers touching this site has to add up to $\Nc$ due to the Grassmann constraint.

This system describes pions that are strongly interacting at low temperatures, resulting in a large entropy, and non-interacting at high temperatures. The chiral transition is located where the pion correlation is maximal (and diverges in the chiral limit). The temperature (in lattice units) is given by $aT=\frac{\gamma^2}{N_t}$ where $\gamma$ is an anisotropy that favors temporal pion hops over spatial hops for $\gamma>1$, and vice versa for $\gamma<1$.

 As explained detail in~\cite{Kim:2023sie}, the integer-valued dual variables can be mapped to a binary vector $x$ which can be adapted for the quadratic unconstrained binary optimization (QUBO) formalism: 
\begin{align}
\label{eq:qubo}
    \chi^2&=x^{T} Q x\,, \nonumber \\
    Q &= W + p(A^{T}A + diag(2 b^T \cdot A) )\,.
\end{align}
Here $Q$ is the quadratic QUBO matrix generated by the weight matrix $W$ that is determined from the effective action, as well as $A$ that incorporates the constraints that stem from the Grassmann integration of the quarks. The penalty factor $p$ favors ($p \gg 1$) or disfavors ($p \ll 1$) the constraint over the weight matrix.
As we explain in more detail in the next section, we embed in parallel $2 \times 2$ sub-lattices such that for each sub-lattice different boundaries $b$ are applied. More precisely, the weight matrix of $Q$ is determined for $\U(3)$ gauge group via 
\begin{align}
    M &=\left(
        \begin{array}{cc}
        -2\log(2m_q)+\log(2) & \log(3) \\
        0 & -\log(2m_q)       \\ 
        \end{array} \right)\\
    D_s&=\left(
        \begin{array}{cc}
        \log(12) &  0 \\
        0 & \log(3) \\        
        \end{array} \right)\\
    D_t&=\left(
        \begin{array}{cc}
        \log(12)-4\log(\gamma) & 0\\
        0 & \log(3)-2\log(\gamma)]\\
        \end{array} \right)\\
    W&=diag(M,M,M,M,D_s,D_s,D_t,D_t) 
    \end{align}
    and the constraint and boundary are given by 
    \begin{align}\label{eq:constraint}
        A&=\left(
        \begin{array}{cccccccccccccccc}
        2 & 1 & 0 & 0 & 0 & 0 & 0 & 0 & 2 & 1 & 0 & 0 & 2 & 1 & 0 & 0 \\
        0 & 0 & 2 & 1 & 0 & 0 & 0 & 0 & 2 & 1 & 0 & 0 & 0 & 0 & 2 & 1 \\
        0 & 0 & 0 & 0 & 2 & 1 & 0 & 0 & 0 & 0 & 2 & 1 & 2 & 1 & 0 & 0 \\
        0 & 0 & 0 & 0 & 0 & 0 & 2 & 1 & 0 & 0 & 2 & 1 & 0 & 0 & 2 & 1 \\
        \end{array} \right)\\
    b&=(3 - d_{ext}^{(1)}, 3 - d_{ext}^{(2)}, 3 - d_{ext}^{(3)}, 3 - d_{ext}^{(4)})\,.
    \label{eq:bc}
\end{align}
The boundary $b$ is no longer constant as in our previous work, but now depends on the number of external dimers $d_{ext}^{(x)}\in \{0,1,2,3\}$.
This quantity changes after each update as explained in Sec.~\ref{sec:sampling}.

\section{Technical development \label{sec:tech}}
In order to sample our effective theory on D-Wave, we have to map logical qubit $x$ as in Eq.~\eqref{eq:qubo} to the physical qubits with connectivities given by the Pegasus topology.
This can be achieved by synchronized combination of physical qubits ("chain") to form a logical qubit, as discussed in detail in \cite{Kim:2023sie}. With 5760 physical qubits available, only small volumes can be sampled at once.
\begin{figure*}[h]
    \centering
    \subfigure[Pegasus Topology]{
    \includegraphics[width=0.48\linewidth]{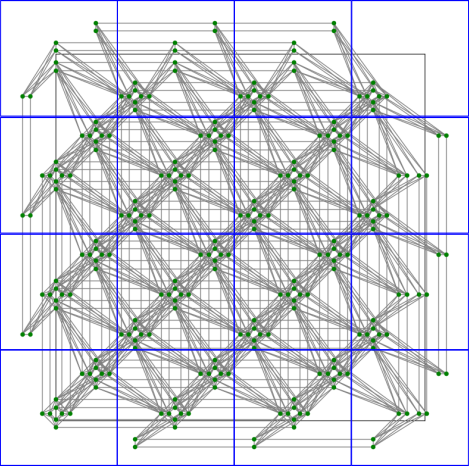}
    }
    \subfigure[Embedding of 4 sub-lattices]{
    \includegraphics[width=0.48\linewidth]{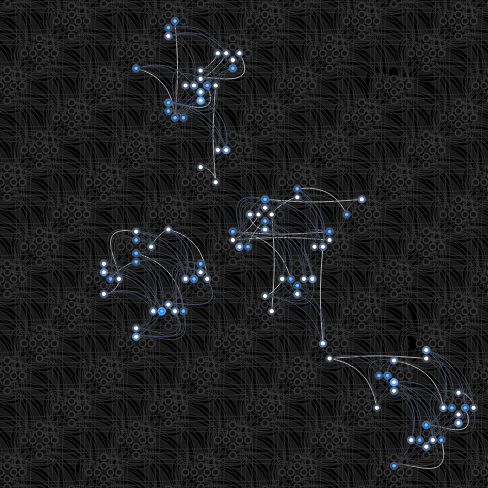}
    }
    \caption{\emph{Left:} the connectivity between qubits of the Pegasus topology, showing a part of the full 5760 qubits. Qubits in this topology are considered to have a nominal length of 12 (each qubit is connected to 12 orthogonal qubits through internal couplers) and degree of 15 (each qubit is coupled to 15 different qubits)\footnote{\url{https://docs.dwavesys.com/docs/latest/c_gs_4.html}}.
 \emph{Right:} embedding of sub-lattices in parallel, here shown for a subset of 4, using automatic embedding, illustrating that each sub-lattice is disconnected. Each sub-lattice uses 24 physical qubits. Each monomer and dimer needs 2 logical qubits to represent integers  between 0 and 3.}
    \label{fig:topology}
\end{figure*}
For larger volumes, we use an iterative scheme that only samples disconnected sub-lattices, keeping dimers that are not part of the sub-lattice fixed, see Eq.~\eqref{eq:bc}. 
An example of the decomposition to sub-lattices is given in Fig.~\ref{fig:fix_bc}. Only the degrees of freedom inside the red boxes are sampled, and this is done in parallel to maximize the use of qubits.
\begin{figure}[ht]
    \centering
    \includegraphics[width=0.49\textwidth]{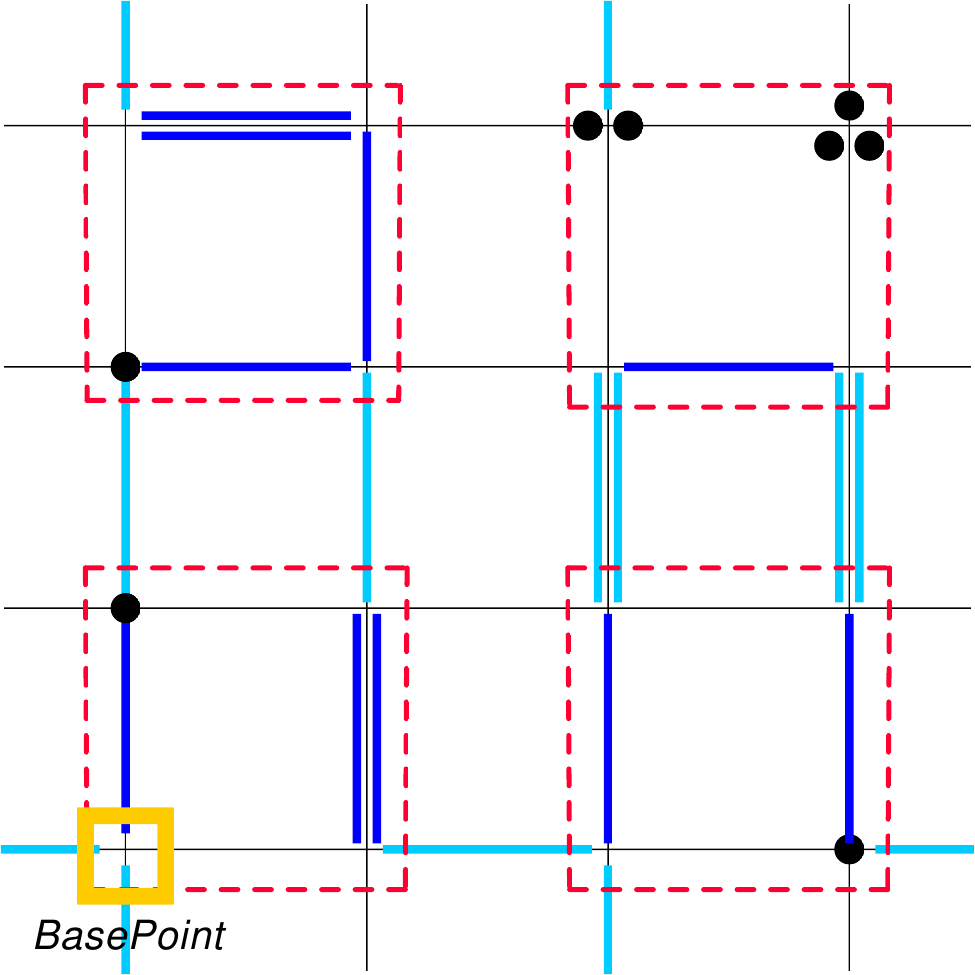}
    \caption{Selecting $2 \times 2$ sub volume with fixed boundary condition. Black dots are monomers, blue lines are dimers. Monomers and dimers inside of red boxes are updated in parallel. The Grassmann constraint is required to be satisfied at each site. The dimers $d_{ext}$ outside of red boxes are external dimers (light blue) and give boundary condition for sampling the red sub-lattices.}
    \label{fig:fix_bc}
\end{figure}

\subsection{Parallel Sampling scheme\label{sec:sampling}}
We now explain the details of the parallel sampling scheme, leaving the details on the Monte Carlo updates for Sec.~\ref{sec:metropolis_hastings}.
In order to sample all degrees of freedom, we have to alternate which dimers are frozen and which are updated. This we do by iterating through the base-points, as high-lighted in Fig.~\ref{fig:fix_bc}, which define what array of sub-lattices have to be considered for the next update. The number of base-points depend on the dimension and the sub-lattice structure, as shown in Table~\ref{tab:parallel}.
\begin{table}[h]
    \centering
    \begin{tabular}{cccc}
        \hline
        V & $V_{\text{sub}}$ & ind. sub-lat. & base points \\
        \hline
        $4 \times 4$ & $2 \times 2$ & 4  & 4 \\
        $8 \times 8$ & $2 \times 2$ & 16  & 4 \\
        $16 \times 16$ & $2 \times 2$ & 64 & 4 \\
        $32 \times 32$ & $2 \times 2$ & 256 & 4 \\
        $4 \times 4 \times 4$ & $2 \times 2$ & 16 & 12 \\
         & $2 \times 2 \times 2$ & 8 & 8 \\
        $4 \times 4 \times 4 \times 4$ & $2 \times 2$ & 64 & 24 \\
         & $2 \times 2\times 2$ & 32 & 32 \\
        $8 \times 8 \times 8 \times 4$ & $2 \times 2$ & 512 & 24 \\
        & $2 \times 2\times 2$ & 256 & 32 \\
        \hline
    \end{tabular}
    \caption{Parallelization by sub-lattice sampling. \#~of ind. sub-lattices corresponds to the number of sub-lattices sampled in parallel. \#~of base points is the number of sites in sub-lattice. For one volume sweep, it is required to sample sub-lattices in parallel for every base point.}
    \label{tab:parallel}
\end{table}
We choose the sub-lattices of size $2 \times 2$ (square sampling), and of size $2 \times 2\times 2$ (cube sampling).
Clearly, the sub-lattice dimension $d_s$ has to be smaller or equal to the dimension $d$ of the lattice volume. For $d_s < d$, we have to loop through $\binom{d}{d_s}$ sets of directions to sample the full volume.
For $2 \times 2$ sub-lattices, 16 logical qubits($\sim 24$ physical qubits) are required per sub-lattice and for $2 \times 2 \times 2$ sub-lattice, 40 logical qubits($\sim 88$ physical qubits). Since the number of sub-lattices is $V/V_{sub}$, the number of logical qubits for lattice size \linebreak 
$V=N_{s}^{(d-1)} \times N_t$ is $N_{\rm logical}=2V(1+ d/2 )$,
with $d$ the dimension of the volume and $N_s$ and $N_t$ are spatial and temporal size of lattice, respectively. Due to the block-diagonal structure of the QUBO matrix $Q$, the embedding of $N_{\rm logical}$ logical qubits to the physical qubits results in disconnected sub-lattice embeddings as shown in Fig.~\ref{fig:topology} (right). The quantum annealer then solves for $x$ in a fixed amount of time regardless of how many qubits are used in the parallelization. 
Hence, as long as the lattice volume $V$ fits on the annealer at once 
(for $2 \times 2$: 170 sub-lattices, $2 \times 2 \times 2$: 45 sub-lattices)
the scaling is constant. If the lattice is too large, we sequentially update a block of maximal subset of sub-lattice at a time. In that case, sampling the sub-lattices scales linear with the volume.

\begin{figure}[h]
\includegraphics[width=1.2\linewidth]{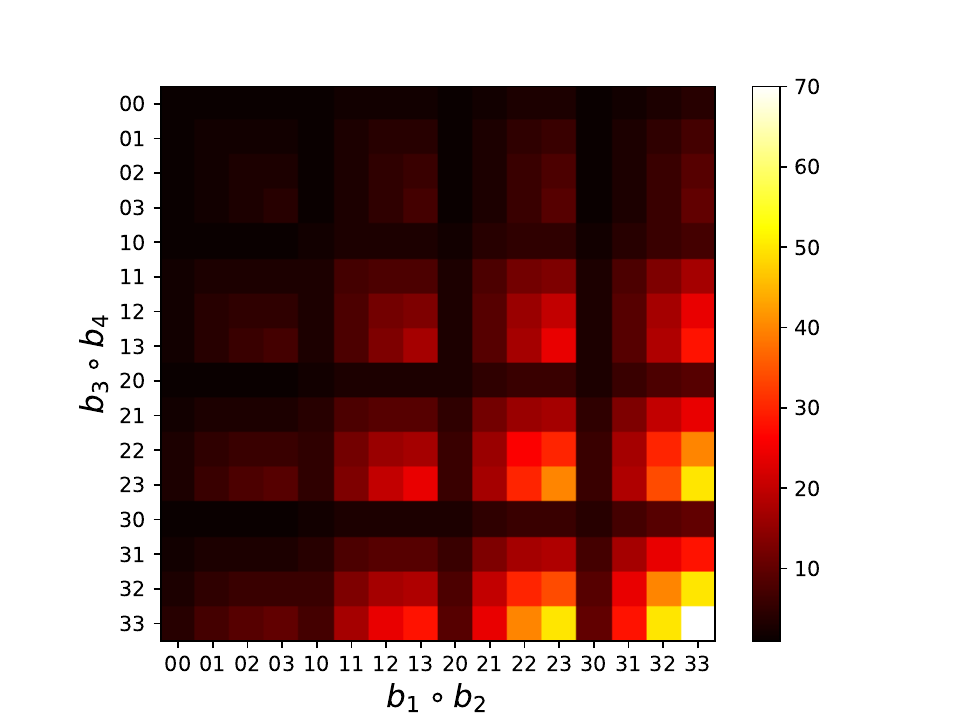}
\caption{Number of configurations for each boundary condition for the square sampling $(b_1, b_2,b_3,b4)$. The color coding reflects the multiplicity of configurations per boundary. For details see App.~\ref{app:ExactEnumeration}} 
\label{fig:heatmap}
\end{figure}

It is sufficient to sample every degree of freedom by sequentially going through all base points that define the location of sub-lattice. The number of base points on a $d$-dimensional lattice is obtained by the binomial coefficient
\begin{align}
    \text{\# of base points}=V_{sub} \times  \binom{d}{d_s}\,.
\end{align}
The solution vector for a new base point depends on the previous solution vector because the fixed boundary $b$ ($d^{(x)}_{ext}$) in Eq.~\eqref{eq:bc} changes after each update. 
For the square sampling, we have $(\Nc+1)^4$ (256 for $\Nc=3$) and for cube sampling we have $(\Nc+1)^8$ (65536 for $\Nc=3$) distinct sets of $b$.
While sampling, we store all generated configurations into histograms. This is feasible for the square update (see Fig. \ref{fig:heatmap}), but for the cube update the larger number of boundaries precludes us.

The quantum annealer produces a distribution of valid configurations that depend on the penalty factor $p$ of the QUBO matrix defined in Eq.~\eqref{eq:qubo}, as well as hyper parameters that we discuss in detail in the next section.
We classify the valid solution vectors per sub-lattice into histograms for each set of boundary conditions $b$, and sort them by monomer number $M$ and number of temporal dimers $D_t$, which encode the quark mass and temperature dependence.
As the set of generated histograms $h_p(b,M,Dt)$ will depend on both the physical parameters and in particular the penalty factor $p$, the histogram is an estimate of the true distribution $h_{true}(b,M,Dt)$ that we obtain from exact enumeration (see Fig.~\ref{fig:heatmap} and App.~\ref{app:ExactEnumeration}. We will discuss in Sec.~\ref{sec:metropolis_hastings} and Sec.~\ref{sec:branching} two methods to correct the histograms to obtain valid results. 

\begin{figure}[h]
{\footnotesize
\begin{align*}
\underbrace{
        \left(
        \begin{array}{c}
        \color{red}{x_1^{(1)}} \\
        \hline
        \color{blue}{x_2^{(1)}} \\
        \hline
        \vdots\\
        \hline
        \color{blue}{x_{n-1}^{(1)}} \\
        \hline
        \color{red}{x_n^{(1)}} \\
        \end{array} \right)\,,
        \left(
        \begin{array}{c}
        \color{blue}{x_1^{(2)}} \\
        \hline
        \color{blue}{x_2^{(2)}} \\
        \hline
        \vdots\\
        \hline
        \color{red}{x_{n-1}^{(2)}} \\
        \hline
        \color{red}{x_n^{(2)}} \\
        \end{array} \right)\,,
        \left(
        \begin{array}{c}
        \color{blue}{x_1^{(3)}} \\
        \hline
        \color{blue}{x_2^{(3)}} \\
        \hline
        \vdots\\
        \hline
        \color{red}{x_{n-1}^{(3)}} \\
        \hline
        \color{blue}{x_n^{(3)}} \\
        \end{array} \right)\,,
        \cdots\,,
        \left(
        \begin{array}{c}
        \color{red}{x_1^{(N)}} \\
        \hline
        \color{red}{x_2^{(N)}} \\
        \hline
        \vdots\\
        \hline
        \color{blue}{x_{n-1}^{(N)}} \\
        \hline
        \color{blue}{x_n^{(N)}} \\
        \end{array} \right)}_{\text{$N$ solution vectors measured}}
\end{align*}
}
\caption{Each column corresponds to a solution vector of a block QUBO matrix. The $x_i^{(j)}$ is a solution vector for each sub-lattice. We collect $x_i^{(j)}$ for fixed $i$ from $j=1,\cdots,N$ with $N$ solution vectors, and check the validity (blue: valid, red: invalid), see Eq.~\eqref{eq:validity} \label{fig:sol}}
\end{figure}

To be more specific, we measure the $N$ independent solution vectors $x^{(j)}$ ($j=1,\cdots, N$), blocked in $n$ separate sub-lattices $x^{(j)}_{i}$ ($i=1, \cdots, n$), 
as shown in Fig.~\ref{fig:sol}.
The boundary condition is the same for all $j$ at fixed $i$. Since not all configurations are valid in the sense that they fulfill the constraint Eq.~\eqref{eq:constraint} (red entries), it may happen that solutions vectors have invalid blocks.  $N$ is typically large enough to ensure that for every $j$ there exists several valid sub-lattices $i$ from which we construct the histograms.

\begin{figure*}[htb!]
    \centering
    \includegraphics[width=0.49\textwidth]{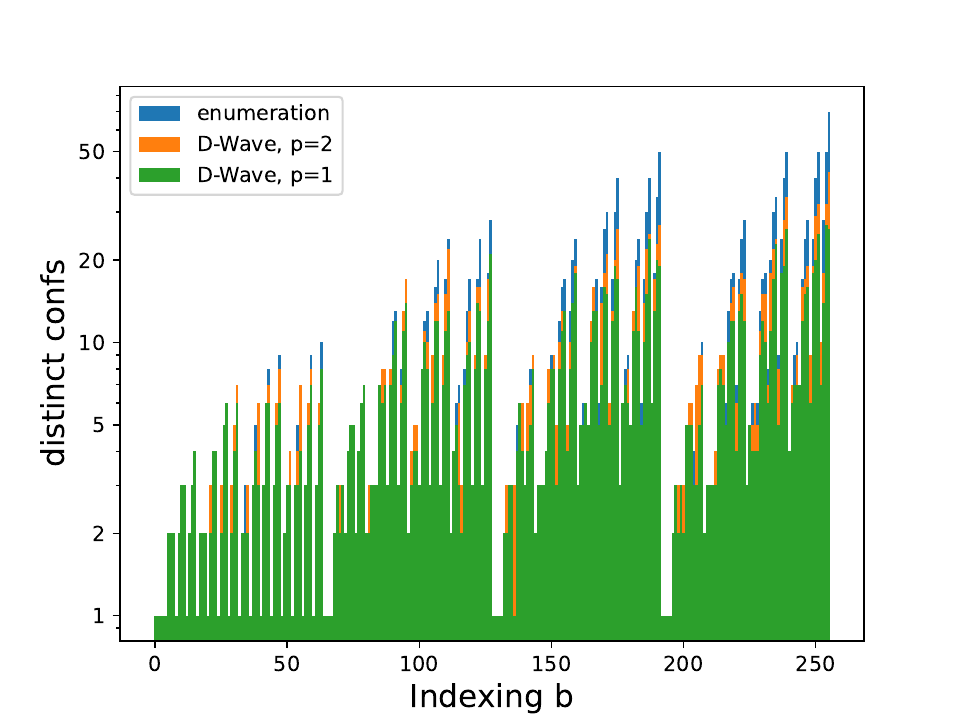}
    \includegraphics[width=0.49\textwidth]{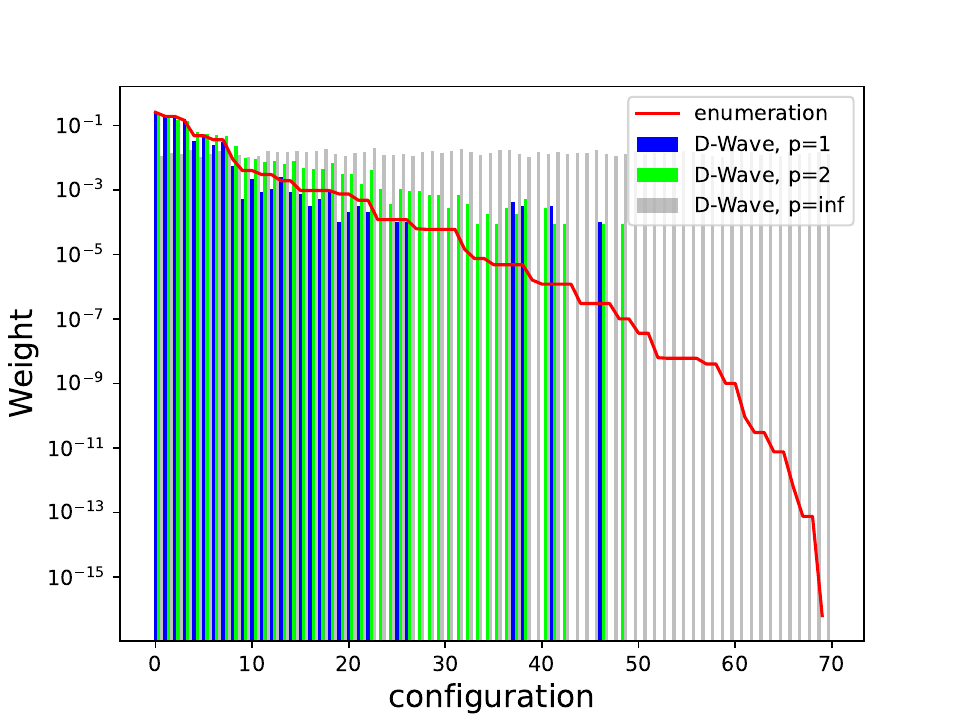}
    \caption{Comparison for $2 \times 2$ sub-lattices between exact histograms from enumeration with those generated by D-Wave for $p=1$ and $p=2$.
		\emph{Left:} Comparison of the multiplicities for 256 different boundaries. \emph{Right:} Comparison of the weights for a specific boundary $b=$(3,3,3,3) for $am_q=1.0$ and $\gamma=0.1$, 
    with exact Boltzmann distribution $h_{true}(b)$ compared to approximate distributions from D-Wave $h_p(b)$. 
    }
    \label{fig:histogram}
\end{figure*}

In Fig.~\ref{fig:histogram}, a comparison of the true histogram 
\begin{align}
h_{true}(b)&=\frac{e^{-S[C_{2\times2}(b)]}}{\sum\limits_{C_{2\times2}(b)} e^{-S[C_{2\times2}(b)]}}
\end{align}
by exact enumeration (with $C_{2\times 2}(b)$ the $2\times 2$ configurations at fixed boundary $b)$ and the histogram $h_p^{b}$ obtained by the D-Wave quantum annealer for penalty factors $p=1, 2, \infty $ is shown. Here, $b$ labels the 256 distinct boundary conditions $b=\{b_1, \dots b_4\}$ in lexicographical order (left). The distribution for a specific boundary condition $b=(3,3,3,3)$ (right) is shown for quark mass $am_q=1.0$ and $\gamma=0.1$, now sorted by the Boltzmann weight $e^{-S}$. Both histograms are in good agreement for $p=1$, but this agreement deteriorates for larger $p$. This indicates that we have good overlap between the true and the approximate distributions of the $2\times2$ sub-lattices, which is important to implement our strategy for large lattice volumes. How to correct the distribution for the small deviations is discussed in Sec.~\ref{sec:metropolis_hastings} and Sec.~\ref{sec:branching}.

\subsection{Optimizing hyper parameters}
To maximize the efficiency of the simulation and obtain correct results, the hyper parameters \texttt{chain\_strength}, \texttt{num\_reads}, \texttt{annealing\_time} should be optimized.
As described earlier, we must embedded logical qubits into the physical layout of qubits so that the constellation of our logical qubits can simulate our QUBO problem (we use D-Wave's automatized embedding to do this mapping).  The physical qubits comprising a logical qubit must act as one;  either all take values of 0 or 1.  And their connections with other logical qubits must also act accordingly.
However, during the annealing process it may (and does) happen that physical qubits within a logical qubit do not act uniformly, and this results in broken ``chains". These broken chains result in invalid solutions.

For submitting a problem to the D-Wave system, one has to provide a value for the \texttt{chain\_strength}, which acts as a constraint that enforces connectivity of the chains. Increasing this value can reduce chain breaks.  However, we cannot arbitrarily increase the value of \texttt{chain\_strength} such that no broken chains occur, since too large a value makes our QUBO matrix less relevant and the system approaches the state of $n$ independent logical qubits that do not interact with each other. This in turn adversely affects the number of obtained valid solutions.  Thus some optimization is required.

\begin{figure*}[htb!]
    \centering
    \subfigure[annealing time=100, \texttt{num\_reads}=800]{
    \includegraphics[width=0.47\textwidth]{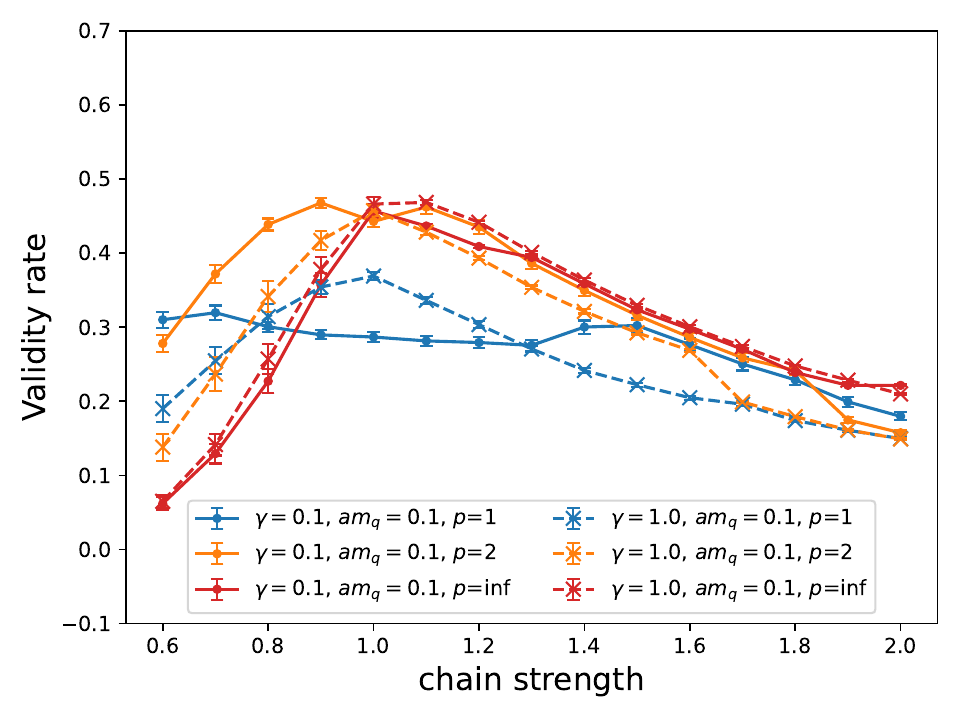}
    }
    \subfigure[annealing time dependence]{
    \includegraphics[width=0.47\textwidth]{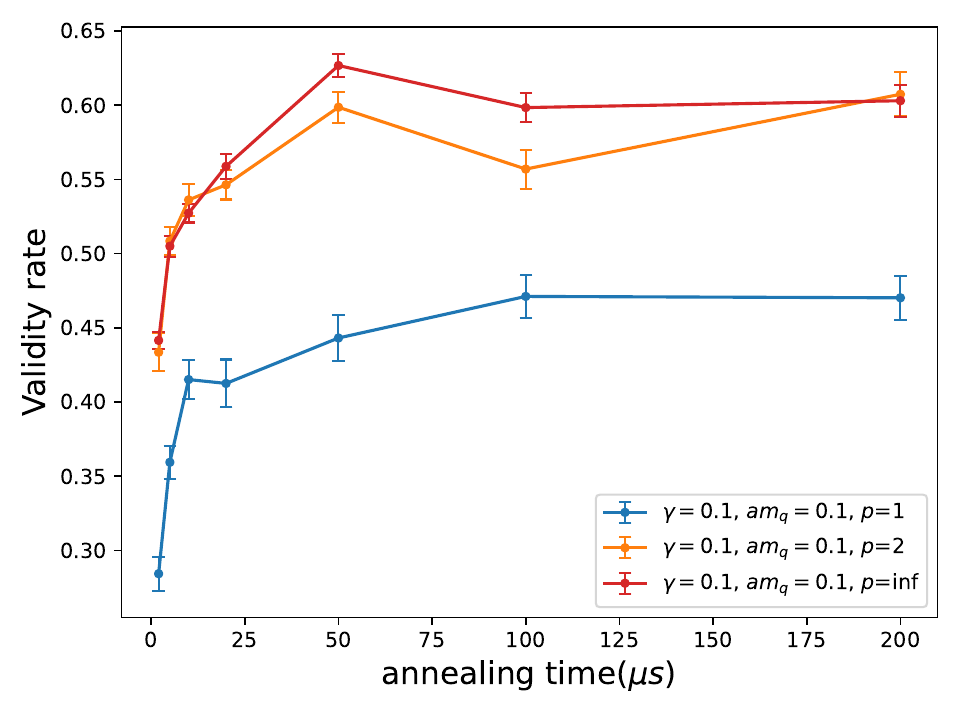}
    }
		\caption{\emph{Left:} The validity rate $v$ as a function of \texttt{chain\_strength} on $2 \times 2$  sub-lattice with '3 3 3 3' boundary condition. For penalty factor $p=1, 2$ and infinite. $\gamma=0.1, 1.0$, $am_q=0.1$. \emph{Right:} The validity rate as function of the annealing time for $p=1, 2$ and infinite and $\gamma=0.1$, $am_q=0.1$. A plateau is reached after about 100ms.
        \label{fig:validity}}
\end{figure*}

We optimize the \texttt{chain\_strength} to maximize the validity rate $v$, which is the number of valid solution vectors over all solution vectors: 
\begin{align}
\label{eq:validity}
v&=\frac{1}{Nn}\sum_{j=1}^{N} \sum_{i=1}^{n} x_{i,\text{valid}}^{(j)} \,.
\end{align}
We rescale the QUBO matrix manually by $Q_{max}$, which is the maximum absolute value of any element of the QUBO matrix, before submitting to the D-Wave solver. This results in the rescaling of the \texttt{chain\_strength} to order one. For small values of \texttt{chain\_strength}, the number of broken chains is large. Since broken chains produce almost always invalid configurations, as a result the validity rate is low, as shown in Fig.~\ref{fig:validity}. With increasing \texttt{chain\_strength}, the validity rate grows with maximum around \texttt{chain\_strength} is between 1.0 to 1.4, depending on the penalty factor $p$ and physical parameters. Note that the weight matrix $W$ in the QUBO matrix has a block-diagonal structure, whereas the constraint matrix $A$ has long-range connectivity, requiring a larger \texttt{chain\_strength}. 
For larger \texttt{chain\_strength}
the validity rate drops again. 

The number $N$ of solution vectors $x$ we request to be generated by the annealer is given by the parameter \texttt{num\_reads}. The duration of an annealing process is set by \texttt{annealing\_time}. A longer anneal time can increase the number of generated valid solutions.  However this is not guaranteed, and the longer anneal times comes at the expense of overall computational cost.
As shown in Fig.~\ref{fig:validity}, we optimize the annealing time per sample. The validity rate is saturated after the annealing time $100 \mu s$. As the annealing time is increased, \texttt{num\_reads} should be decreased accordingly to keep the computational cost constant. 
We determine the optimal \texttt{chain\_strength}/annealing time from the maximum/plateau of the validity rate.
Since we have limited amount of quantum computing time, to balance the quantum computing time consumption and quality of the solution, we choose the \texttt{chain\_strength} 1.2, $\texttt{annealing\_time}=100 \mu s$ and $\texttt{num\_reads}=200$. This allows us to generate histograms for more physical parameters of the interest.

\subsection{Metropolis and Metropolis-Hastings}
\begin{figure*}[htb!]
    \centering
    \subfigure[Metropolis-Hastings (D-Wave) \newline p-dependence]{\includegraphics[width=0.31\linewidth]{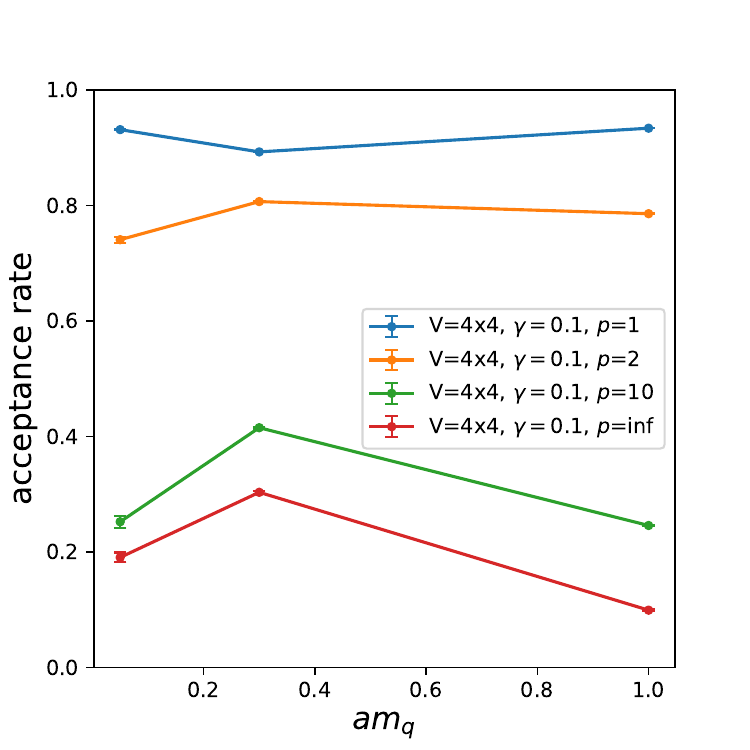}}
    \subfigure[Metropolis-Hastings (D-Wave) \newline volume-dependence]{\includegraphics[width=0.31\linewidth]{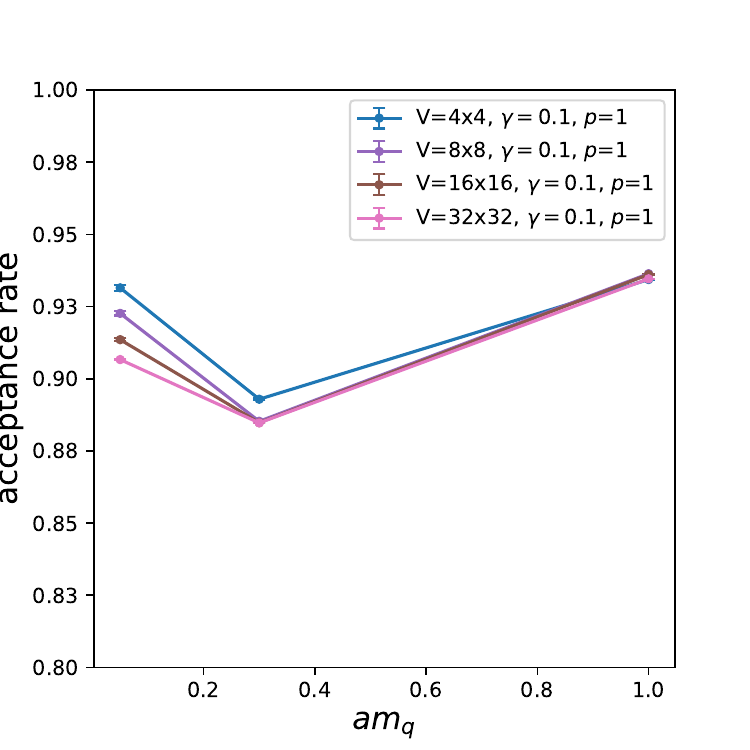}}
    \subfigure[Metropolis (classical)]{\includegraphics[width=0.31\linewidth]{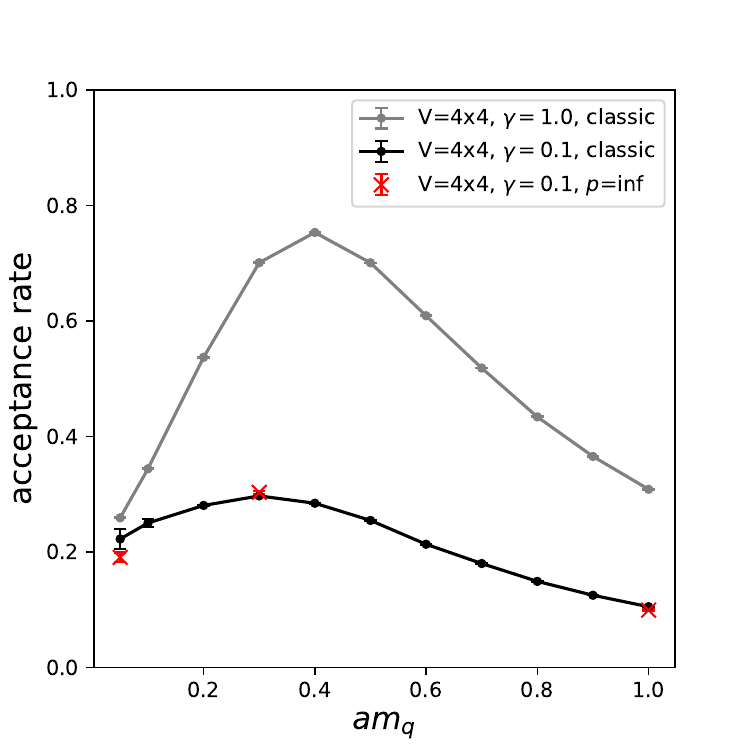}}
    \caption{Comparison of the acceptance rate versus $am_q$. \emph{Left:} for a $4\times4$ lattice from Metropolis-Hastings, with $2 \times2$ sub-lattice histograms from D-Wave, for various penalty factors $p$. \emph{Center:} illustrating independence  of the acceptance rate of the volume. \emph{Right:} Metropolis from classical computation, for $\gamma=0.1, 1.0$. Metropolis-Hastings for $p=\infty$ reproduces the classical Metropolis acceptance rate.}
    \label{fig:accept_Metro-Hastings}
\end{figure*}

{\label{sec:metropolis_hastings}}
Although we know the exact distributions of the $2\times2$ sub-lattices from enumeration in this simplest case (see App.~\ref{app:ExactEnumeration}), this is not the case in the more realistic cases in general (SU(3) including strong coupling expansion). But even with this exact distribution, it is not feasible to extend to finite periodic lattices:
First, analytically extending to larger sub-lattices such as $4\times4$ can not be achieved due the exponential growth of computation involved. Second, a standard Metropolis algorithm attempting to glue the sub-lattices together would produce extremely small acceptance rates of about $0.1\%$ independent of the physical parameters.
Alternative classical algorithms such as the Metropolis algorithm using the parallelization scheme as in Fig.~1 and sweeping through the 4 base points gives acceptance rates of about 20\% for low temperatures $\gamma=0.1$, with a quark mass dependence as shown in Fig.~\ref{fig:accept_Metro-Hastings} (right).

\begin{figure}
    \centering
    \includegraphics[width=\linewidth]{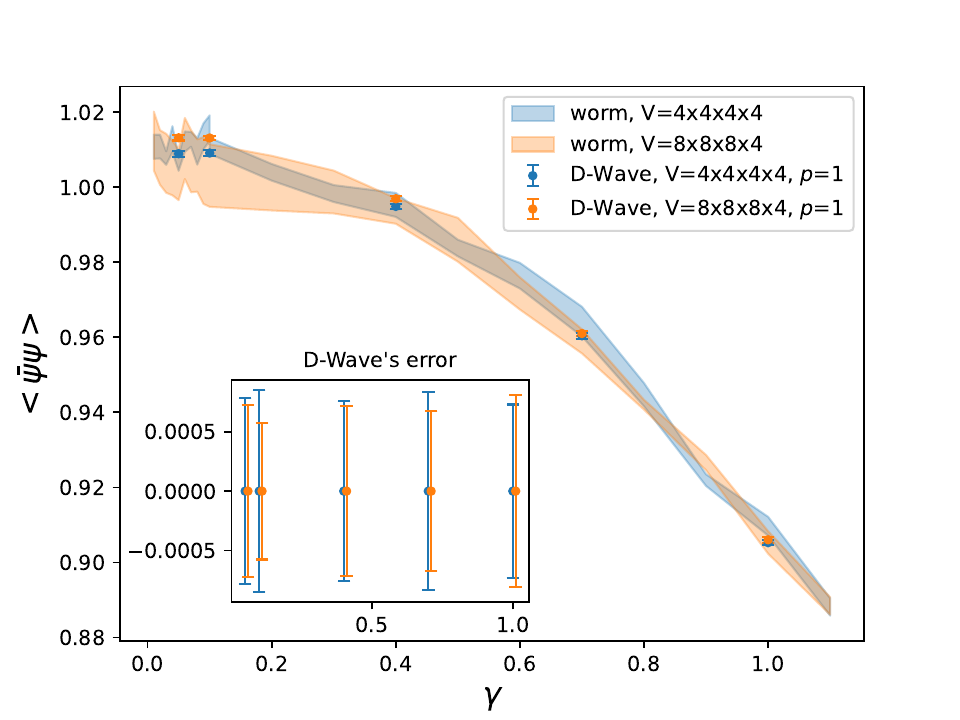}
    \caption{The chiral condensate as a function of $\gamma$ which is related to the temperature $aT=\frac{\gamma^2}{N_T}$. While the size of error of classical simulations via worm algorithm depend on the physical parameter, the method that we demonstrate in this paper is independent of $\gamma$, as illustrated in the inset.  Hence, we confirm that this method has a significant advantage over the classical method at low temperatures.} 
    \label{fig:result_chiral_vs_gamma_4D}
\end{figure}

In contrast, with the approximate distribution for $2\times 2$ sub-lattices measured by D-Wave for penalty factor $p=1$, we do not rely at all on the exact distribution obtained from classical computation at all. However, we cannot incorporate the standard Metropolis algorithm, but have to use the Metropolis-Hastings algorithm: 
\begin{align}
P_{\rm accept}&=e^{-S_{\rm new}+S_{\rm old}}\frac{h_{\rm old}}{h_{\rm new}}
\label{eq:MetropolisHastings}
\end{align}
with $e^{-S_{\rm new}}$ and $e^{-S_{\rm old}}$ the Boltzmann factors for the new/old configuration, and $h_{\rm new}$ and $h_{\rm old}$ the new/old histograms of the configuration. They are non-trivial: although they have the same boundary condition $b$, they differ in the specific distribution of monomers and dimers within the $2\times2$ sub-lattice.
The proposal probability $P_{\rm proposal}=\frac{h_{\rm old}}{h_{\rm new}}$ for a set of $2 \times 2$ updates is not uniform as it is drawn from the measured histogram. The details of the procedure are described by the following steps:
\begin{enumerate}
    \item On the classical computer, generate a large lattice filled with 3 monomers at every sites. 
    \item Select independent sub-lattices and use the set of boundaries $\{b\}$ to construct the QUBO matrix for the physical parameters. 
    \item Generate a distribution for the sub-lattices using the quantum annealer and store it on the memory of classical computer. 
    If the distribution for a specific boundary condition already exists, we can skip this step.
    \item Select for each sub-lattice $h_{new}$ from the distribution and make an accept/reject step by comparing with the previous configuration $h_{old}$ on the classical computer according Eq.~\eqref{eq:MetropolisHastings}. 
    \item Store the configuration in terms of monomers and dimers and measure observables.
    \item Change the base point. Repeat from step 2. 
\end{enumerate}
As pointed out in Sec.~\ref{sec:sampling}, these histograms have good overlap with the true distribution. Hence we expect that the Metropolis-Hastings algorithm will have a much larger acceptance rate, in particular also for low temperatures. The dependence of the acceptance probability on the penalty factors $p$ is shown in Fig.~\ref{fig:accept_Metro-Hastings} (left).
We can clearly see that $p=1$ has the largest acceptance rate, and that large $p$ will decrease the acceptance rate. 
In the limit $p\rightarrow \infty$, the weight matrix will not contribute to the QUBO matrix and the $2 \times 2$ histograms become flat. Hence, due to $h_{old}=h_{new}$, the proposal probability drops out in the Metropolis-Hastings, which simplifies to standard Metropolis. The acceptance rate will hence be minimal as there is poor overlap with the important configurations given by the physical parameters. As shown in Fig.~\ref{fig:accept_Metro-Hastings} (c), the $p=\infty$ acceptance rate agrees with the classical Metropolis algorithm.
Recall that classical computations are in particular expensive for low temperatures (small $\gamma$), whereas the computational costs for the histograms measured by D-Wave do not depend on $\gamma$.  
This is particularly evident by comparing the statistical uncertainties of the chiral condensate as measured via the worm algorithm versus the Metropolis-Hastings results based on the D-Wave histograms, as shown in Fig.~\ref{fig:result_chiral_vs_gamma_4D}. In the case of worm simulation, the statistical uncertainty grows with decreasing temperature, while it remains constant for the results based on the hybrid approach of D-Wave histograms and Metropolis-Hastings. Interestingly, the error for $8^3\times 4$ is of comparable size as for $4^3\times 4$, both having the same number of Metropolis-Hastings updates. This might be related to the fact that for each base point, the solution vector $x$ scales with the volume, as it contains $2\times 2$ sub-lattices proportional to the volume.

Also, the acceptance rate does not depend much on the volume as shown in Fig.~\ref{fig:accept_Metro-Hastings} (b), as expected. 

\begin{figure*}[htb!]
    \centering
    \subfigure[Chiral Condensate at $\gamma=0.1$]{\includegraphics[width=0.48\linewidth]{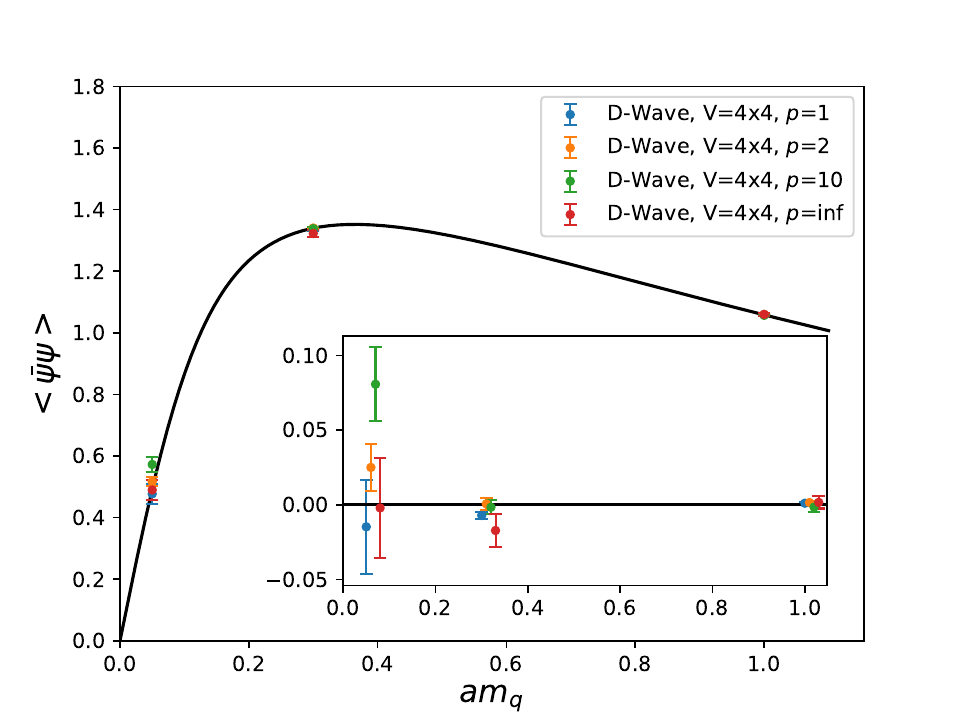}}
    \subfigure[Energy Density at $am_q=0.3$]{\includegraphics[width=0.48\linewidth]{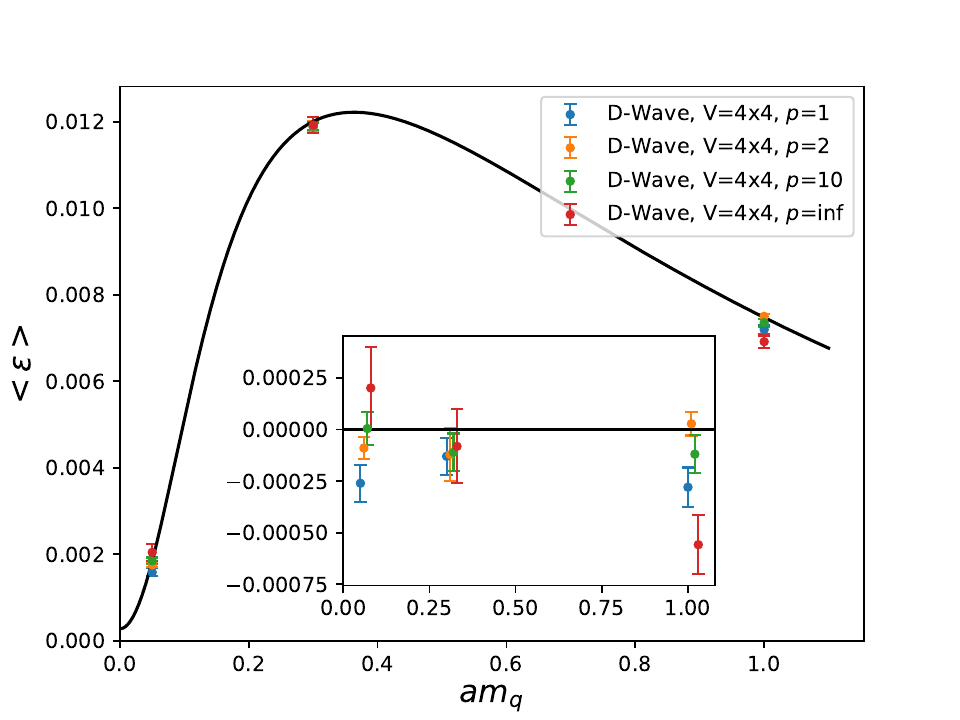}}
    \caption{Results from the Metropolis-Hastings algorithm, all for $\gamma=0.1$: chiral condensate (left) and energy density (right) for different $p$-values. The solid lines are analytic results from an exact enumeration. The inset figures show the difference between analytic and the numerical results. For any quark mass, $p=1$ has the smallest error.}
    \label{fig:observable}
\end{figure*}

\begin{figure*}[htb!]
    \centering
    \subfigure[Chiral Condensate at $\gamma=0.1$]{\includegraphics[width=0.48\linewidth]{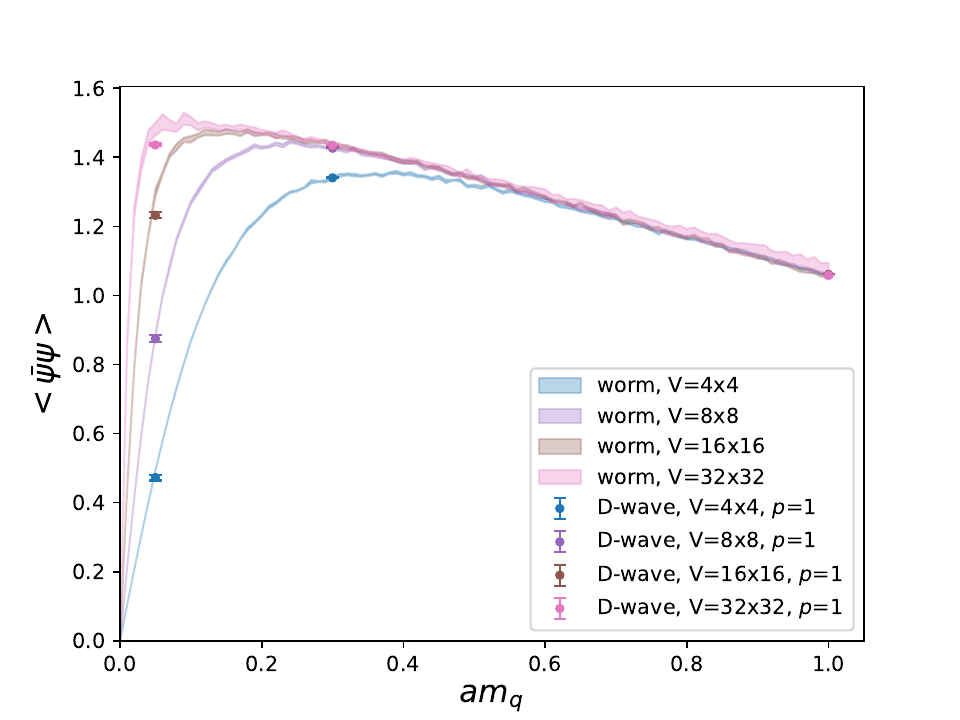}}
    \subfigure[Energy Density at $am_q=0.3$]{\includegraphics[width=0.48\linewidth]{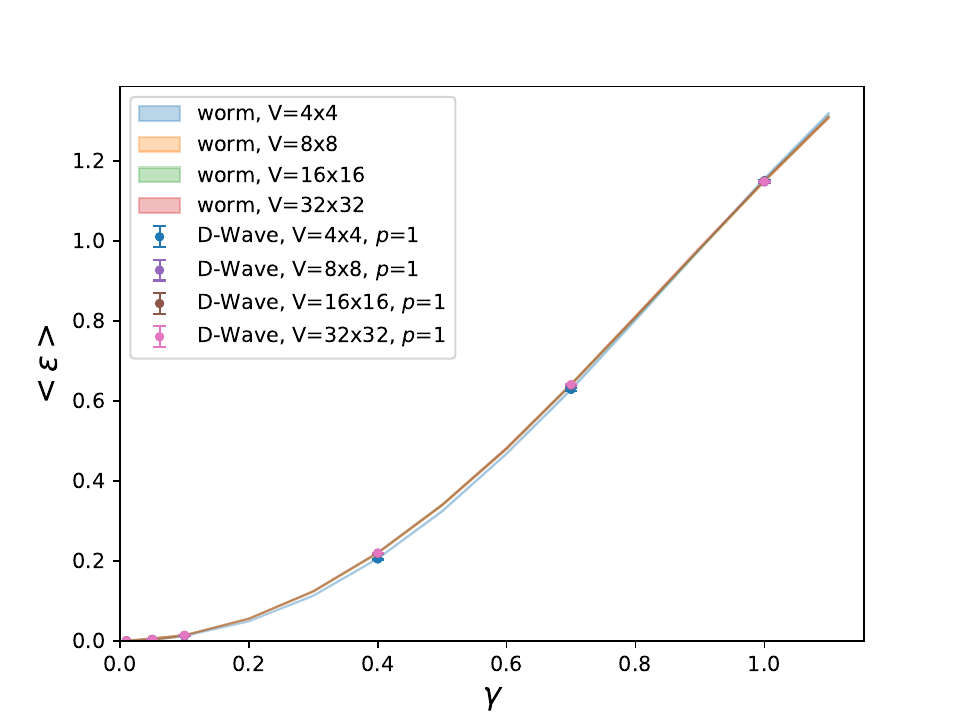}}
    \caption{Results of Metropolis-Hastings method for the chiral condensate for $p=1$ at $\gamma=0.1$ (left) and the energy density (right) for $p=1$ at $am_q=0.3$ for various 2-dimensional volumes.} 
    \label{fig:result_Metro-Hastings_largeV}
 \end{figure*}

For the thermodynamic observables such as the chiral condensate and energy density, we show in Fig.~\ref{fig:observable} the $p$-dependence for $p=1,2,10,\infty$. We clearly see in the inset of Fig.~\ref{fig:observable} (a) that $p=1$ results in the smallest error, as it has the largest acceptance rate. For the large volume results in the next section, we hence always use $p=1$.
We show in Fig.~\ref{fig:result_Metro-Hastings_largeV} 
for $p=1$ the volume-dependence of the chiral condensate and the energy density,
for $V=4\times4, 8\times8, 16\times16, 32\times32$ and compare the data with results from a classical computer using the worm algorithm.

\subsection{Branching Strategy}
\label{sec:branching}
In the previous section we studied in detail how to obtain the correct equilibrium distribution from the histograms that D-Wave provides via the Metropolis-Hastings algorithm. Note that if the histograms were already perfect Boltzmann distribution, we could instead use a heatbath algorithm.
Another promising method to obtain valid results 
that we briefly outline is to repeatedly branch into sub-branches, and statistically evaluate each branch to obtain expectation values for thermal observables, such as the chiral condensate and the energy. The essential idea of the branching strategy is to maximize quantum parallelization. While in the Metropolis-Hastings the same accept-reject step is required, the branching strategy performs accept-reject at each branch. 

We start with a specific boundary condition (square sampling) with $b=(3,3,3,3)$, i.e. we choose $2 \times 2$ sub-lattices with weights depending on the physical parameters $\gamma$ and $am_q$, as shown in Fig.~\ref{fig:histogram} (right). The number of samples $N_{\rm branch}$ defines the branches. The same sample can be chosen multiple times in extreme cases (e.g. $am_q$ very large), although this is rare for physical parameters of interest (low temperatures, intermediate and small quark masses). 

If we restrict to 2 dimensions, as we have 4 base points for $2 \times 2$ sub-lattices, we have to repeat the above procedure 4 times and obtain further sub-branches, in total $N_{\rm branch}^4$ branches, each representing a valid configuration. 
While branching into $i$ sub-branches, the weight factor $w_i=e^{-S[i]}/h_{p,i}^b$ for each sample with index of depth $i$ multiplies the weight factors for all existing branches, where $h_p^b$ is the entry of the measured histogram, with $b$ the specific boundary condition. The final weight for each of the $N_{\rm branch}^4$ branches is
\begin{align}
W_{\rm branch}=w_1 \times w_2 \times w_3 \times w_4
\end{align}
Ultimately, to correct for bias, we must reweight our observable $\mathcal{O}$, 
\begin{displaymath}
\langle O W_{\rm branch}\rangle/ \langle W_{\rm branch}\rangle\ .
\end{displaymath}
Unfortunately a limitation of this method so far is that the memory requirement for branching grows exponentially with $N_{\rm branch}$, preventing us from presenting results at this point in time.

\section{Results\label{sec:results}}
\begin{figure*}[htb!]
\subfigure[Chiral Condensate at $\gamma=0.1$]{
    \includegraphics[width=0.48\textwidth]{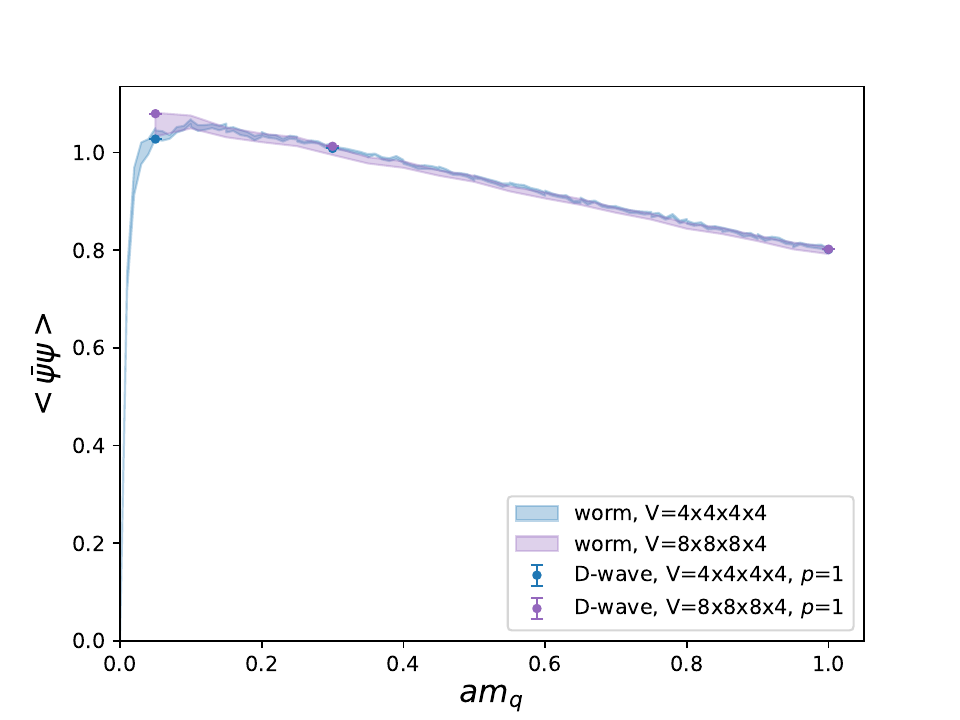}
    }
    \subfigure[Energy Density at $am_q=0.3$]{
    \includegraphics[width=0.48\textwidth]{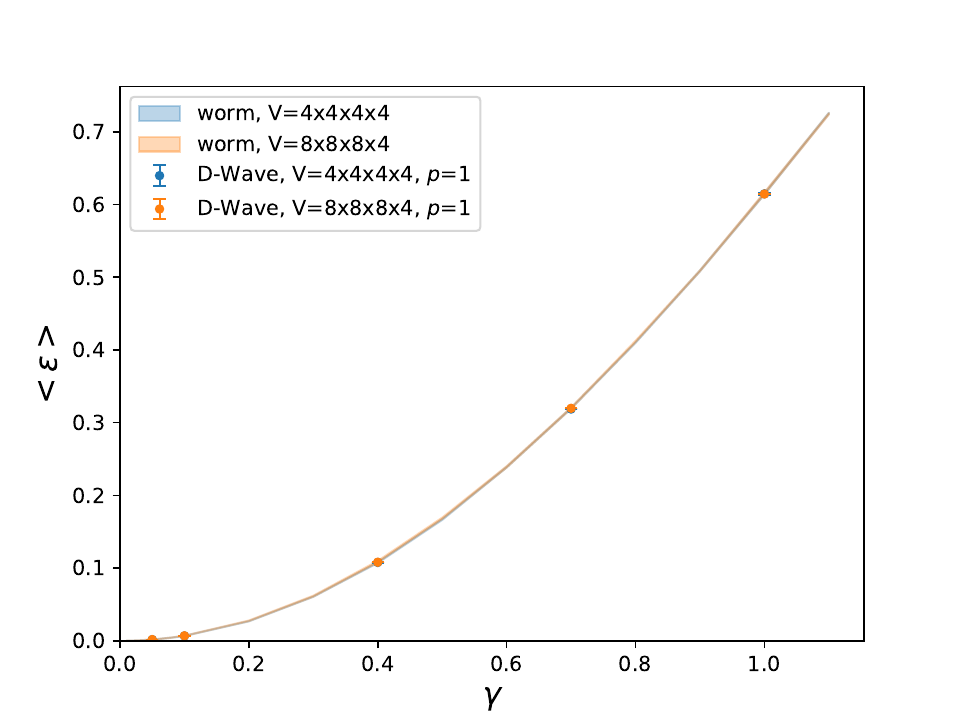}
    }
		\caption{4-dim. results on $V=4 \times 4 \times 4 \times 4$ and $8 \times 8 \times 8 \times 4$. \emph{Left:} chiral condensate at $\gamma=0.1$ for $am_q=0.05,0.3,1.0$. \emph{Right:} energy density at $am_q=0.3$ for $\gamma=0.05,0.1,0.4,0.7,1.0$. \label{fig:results_4d}}
\end{figure*}

\begin{figure*}[hbt!]
\subfigure[Chiral extrapolation at $\gamma=0.1$] {\includegraphics[width=0.48\linewidth]{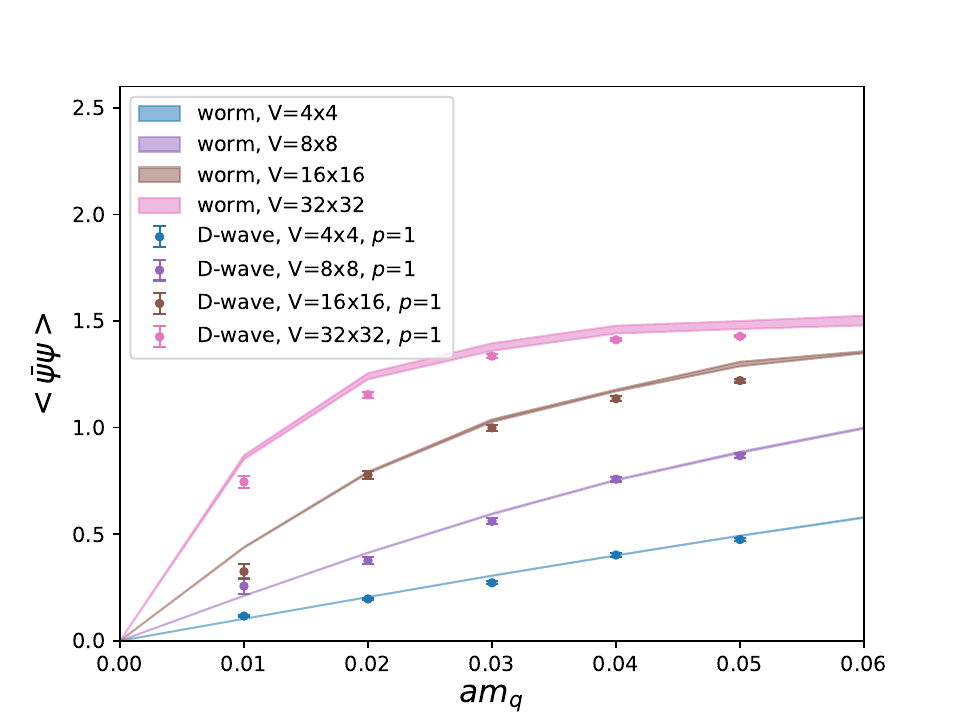}}
\subfigure[Low temperature extrapolation at $am_q=0.3$] {\includegraphics[width=0.48\linewidth]{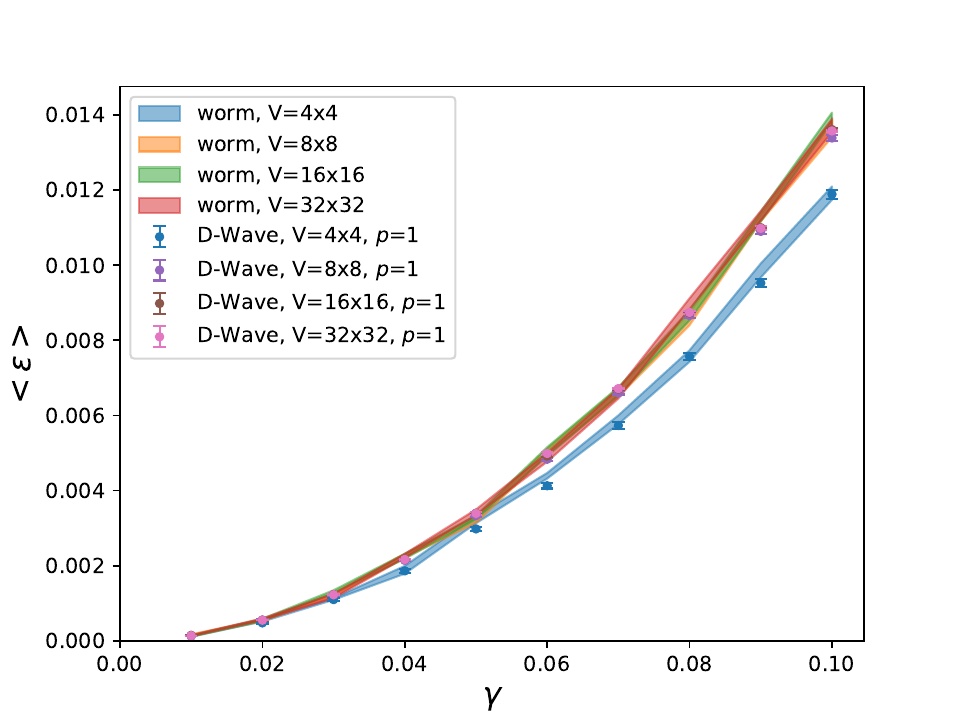}}
   \caption{Chiral extrapolation (left) from histograms at $am_q=0.05$, $\gamma=0.1$, $p=1$, obtained via Metropolis-Hastings for several quark masses $am_q \leq 0.05$ and various volumes. Right: low temperature extrapolation from $am_q=0.3$, $\gamma=0.1$, $p=1$ for several $\gamma \leq 0.1$. The bands are obtained by worm simulation on the classical computer.} 
    \label{fig:result_extrap}
\end{figure*}

We have compared different 2-dimensional lattices in Sec.~\ref{sec:metropolis_hastings} and extended to 4-dimensional volumes for the thermodynamic observables, as shown in Fig.~\ref{fig:results_4d},
which includes a $4^3\times 4$ and $8^3\times 4$ lattice.
We are particularly interested in the chiral limit $am_q\rightarrow 0$ for the chiral condensate, and the low temperature limit $\gamma\rightarrow 0$ for the energy density, as these limits are particularly interesting from the physics point of view, and expensive with classical computers. We also present  ``extrapolated results", for which we use a histogram at fixed quark mass and temperature $\gamma$ and run the Metropolis-Hastings for smaller quark masses and/or temperatures, given that there is still sufficient overlap with the corresponding target distributions. 
We demonstrated that these extrapolations are under control for all 2-dimensional volumes under investigation: the lattices $4\times 4$, $8\times 8$, $16\times 16$, $32\times 32$, as shown in Fig.~\ref{fig:result_extrap}.
As this works remarkably well, we can reduce drastically the number of histograms that have to be generated on the annealer for the set of physical parameters. 

\section{Conclusion}

We developed an algorithm to simulate $\U(3)$ gauge theory in the strong coupling limit for large 2-dimensional and 4-dimensional volumes on a quantum annealer. In particular, we made use of an hybrid quantum/classical approach, whereby we determined the histograms in the dual variables relevant to the QUBO formalism via the D-Wave quantum annealer, and than ran the Metropolis-Hastings algorithm based on these histograms on a classical computer. 
We described the hyper parameters we used to reduce the compute time on the annealer, detailing how we numerically determined optimal values for \texttt{annealing\_time} and \texttt{chain\_strength} by maximizing the validity rate given our finite compute time resources. Also, the penalty factor $p$ that balances the weight matrix and the constraint in the QUBO matrix, for which the quality of the histograms is best was determined to be $p=1$, resulting in large acceptance rates.

Further improvements can be obtained by using the quantum parallelization more extensively and efficiently while branching our sub-lattices. This would require a better understanding of how the number of branches can be optimized with respect to compute time and memory requirement, in particular in four dimension. We are actively investigating this option.

Extending to $\SU(3)$ and including gauge corrections are another important next step towards a more realistic effective theory of lattice QCD. The first aspect includes baryons while the second aspect includes gluon propagation. The dual representation is well established for both extensions, and again can be mapped on binary vectors. However, these formulations will require more logical qubits. Still, it remains feasible to map the logical qubits required for the QUBO matrix to the physical qubits on the annealer.
However, while the histograms required for this work, given square sampling, had $4^4$ different boundaries to classify the histograms, these generalizations will require up to $16^4$ different boundaries, for which the histograms can no longer be pre-computed on D-Wave. Instead they need to be computed during Metropolis-Hastings in between the updates, whenever required. 
Studying more realistic effective theories will introduce two more physical parameters~\cite{Kim:2023dnq}: the quark chemical potential $\mu_q$, and the inverse gauge coupling $\beta$ that is related to the lattice spacing $a$.
We will report on our findings using the same approach via histograms determined on the annealer in a subsequent publication. 

\begin{acknowledgments}
The authors gratefully acknowledge the J\"ulich Supercomputing Centre (https://www.fz-juelich.de/ias/jsc) for funding this project by providing computing time on the D-Wave Advantage™ System JUPSI through the J\"ulich UNified Infrastructure for Quantum computing (JUNIQ).
J.K. and T. L. were supported by the Deutsche Forschungsgemeinschaft (DFG, German Research Foundation) through the funds provided to the Sino-German Collaborative Research Center TRR110 "Symmetries and the Emergence of Structure in QCD" (DFG Project-ID 196253076 - TRR 110) and as part of the CRC 1639 NuMeriQS–project no. 511713970, respectively.
J.K. and W.U. are supported by the Deutsche Forschungsgemeinschaft (DFG) through the CRC-TR 211 'Strong-interaction matter under extreme conditions'– project number 315477589 – TRR 211. 
\end{acknowledgments}

\appendix

\section{Exact Enumeration}
\label{app:ExactEnumeration}

From exact enumeration, we obtain for $2 \times 2$ sub-lattices with fixed boundary conditions (the building blocks of square sampling) in total 2350 distinct configurations, that are distributed over 256 distinct boundaries $(b_1,b_2,b_3,b_4)$, with $b_i\in \{0,1,2,3\}$.

If we further distinguish the sub-lattice configurations by its monomer number $M$ and number of temporal dimers $D_t$, we obtain refined multiplicities. 
In the Tab.~\ref{tab:enumeration} we give the multiplicities in these sectors $M$, $D_t$.

\begin{table}[htb!]
\begin{tabular}{rrrrrrrrr}
\hline
$M$ &  &  &  & $D_t$ &  &  &  &  sum \\
 & 0 &1 &2 &3 &4 &5 &6 & \\
\hline
0 & 16 & 18 & 17 & 10 & 6 & 2 & 1 &  70\\
1 & 48 & 60 & 48 & 28 & 12 & 4 & 0 & 200\\
2 & 92 & 110 & 82 & 40 & 14 & 2 & 0 & 340\\
3 & 132 & 148 & 96 & 40 & 8 & 0 & 0 & 424\\
4 & 153 & 154 & 88 & 26 & 3 & 0 & 0 & 424\\
5 & 148 & 132 & 60 & 12 & 0 & 0 & 0 & 352\\
6 & 124 & 92 & 32 & 4 & 0 & 0 & 0 & 252\\
7 & 88 & 52 & 12 & 0 & 0 & 0 & 0 & 152\\
8 & 54 & 24 & 3 & 0 & 0 & 0 & 0 & 81\\
9 & 28 & 8 & 0 & 0 & 0 & 0 & 0 & 36\\
10 & 12 & 2 & 0 & 0 & 0 & 0 & 0 & 14\\
11 & 4 & 0 & 0 & 0 & 0 & 0 & 0 & 4\\
12 & 1 & 0 & 0 & 0 & 0 & 0 & 0 & 1\\
\hline
sum & 900 & 800 & 438 & 160 & 43 & 8 & 1 & 2350\\
\hline
\end{tabular}
\caption{Multiplicities from exact enumeration on $2 \times 2$ sub-lattices with fixed boundary conditions sorted by monomer number $M$ and temporal dimer $D_t$ }
\label{tab:enumeration}
\end{table}

\begin{table*}[htb!]
\begin{tabular}{rrrrrrrrrrrrrrr}
\hline
$M$ &  &  & & & & & & $B$ &  &  &  & & & sum \\
 & 0 &1 &2 &3 &4 &5 &6 & 7 & 8 & 9 & 10 & 11 & 12& \\
\hline
0 & 1 & 0 & 4 & 0 & 10 & 0 & 20 & 0 & 19 & 0 & 12 & 0 & 4 & 70\\
1 & 0 & 4 & 0 & 16 & 0 & 40 & 0 & 64 & 0 & 52 & 0 & 24 & 0 & 200\\
2 & 0 & 0 & 10 & 0 & 40 & 0 & 88 & 0 & 116 & 0 & 74 & 0 & 12 & 340\\
3 & 0 & 0 & 0 & 20 & 0 & 72 & 0 & 136 & 0 & 144 & 0 & 52 & 0 & 424\\
4 & 0 & 0 & 0 & 0 & 31 & 0 & 100 & 0 & 158 & 0 & 116 & 0 & 19 & 424\\
5 & 0 & 0 & 0 & 0 & 0 & 40 & 0 & 112 & 0 & 136 & 0 & 64 & 0 & 352\\
6 & 0 & 0 & 0 & 0 & 0 & 0 & 44 & 0 & 100 & 0 & 88 & 0 & 20 &252\\
7 & 0 & 0 & 0 & 0 & 0 & 0 & 0 & 40 & 0 & 72 & 0 & 40 & 0 &152\\
8 & 0 & 0 & 0 & 0 & 0 & 0 & 0 & 0 & 31 & 0 & 40 & 0 & 10 &81\\
9 & 0 & 0 & 0 & 0 & 0 & 0 & 0 & 0 & 0 & 20 & 0 & 16 & 0 & 36\\
10 & 0 & 0 & 0 & 0 & 0 & 0 & 0 & 0 & 0 & 0 & 10 & 0 & 4 & 14\\
11 & 0 & 0 & 0 & 0 & 0 & 0 & 0 & 0 & 0 & 0 & 0 & 4 & 0 & 4 \\
12 & 0 & 0 & 0 & 0 & 0 & 0 & 0 & 0 & 0 & 0 & 0 & 0 & 1 & 1 \\
\hline
sum & 1 & 4 & 14 & 36 & 81 & 152 & 252 & 352 & 424 & 424 & 340 & 200 &  70 & 2350\\
\hline
\end{tabular}
\caption{Multiplicities from exact enumeration on $2 \times 2$ sub-lattices with fixed boundary conditions sorted by monomer number $M$ and sum of boundary conditions $B=b_0+b_1+b_2+b_3$}
\label{tab:enumeration2}
\end{table*}

\bibliography{references}

\end{document}